\documentclass{mhd}     

\usepackage{graphicx}   	
\usepackage{amsmath}

\mhdhead{0}{0}{1}	

\title{Influence of Ohmic and ambipolar heating on thermal structure of accretion discs}	
\author{S~.A.~Khaibrakhmanov\inst{1}\inst{,2}, \and A.~E.~Dudorov\inst{2}} 	
\institute{Ural Federal University, 51 Lenin str., Ekaterinburg 620000, Russia
\and
Chelyabinsk state university, 129 Br. Kashirinykh str., Chelyabinsk 454001, Russia} 
		
\begin{document}

\maketitle
\begin{abstract}%
We investigate dynamics of accretion discs of young stars with fossil large-scale magnetic field. Our magneto-gas-dynamic (MHD) model of the accretion discs includes equations of Shakura and Sunyaev, induction equation, equations of thermal and collisional ionization. Induction equation takes into account Ohmic and magnetic ambipolar diffusion, magnetic buoyancy.

We also consider the influence of Ohmic and ambipolar heating on thermal structure of the accretion discs. We analyse the influence of considered dissipative MHD effects on the temperature of the accretion discs around classical T Tauri star. The simulations show that Ohmic and ambipolar heating operate near the borders of the region with low ionization fraction (`dead' zone). Temperature grows by $\sim 1000$~K near the inner boundary of the `dead' zone, $r\approx (0.5-1)$~au, and by $\sim 100$~K near its outer boundary, $r\approx (30-50)$~au.
\end{abstract}

\section*{Introduction.}
Observations in optical, infrared (IR), submillimeter and radio ranges have shown that young stars are surrounded by gas-dust discs~\cite{williams11}. The discs are geometrically thin and optically thick. Typical masses of the discs are $0.001-0.1M_{\odot}$, radii are 100-1000~au\footnote{$1\,\mathrm{AU} = 1.5\times 10^{13}$~cm}. Matter accretes from the disc onto the star with mass rates $\dot{M}=(10^{-9}-10^{-6})\,M_{\odot}\,\mathrm{yr}^{-1}$ over time $10^6-10^7$~yr.

It has been found that thermal emission of dust grains in the accretion discs of young stars (ADYS) is polarized \cite{stephens14, li16, li18}. The polarization can be caused by paramagnetic dust grain alignment by the magnetic field in the disc. Therefore, the polarimetric studies point to existence of large scale magnetic field in the accretion discs. Measurements of the magnetic field strength in the ADYS are still challenging. 

Young stars with accretion discs form as a result of gravitational collapse of magnetized cores of molecular clouds. Numerical simulations have shown that initial magnetic flux of the cloud cores partially conserves during the star formation (see reviews~\cite{inutsuka12, fmft}). Therefore it is natural to assume that the large-scale magnetic field of the accretion disc has fossil nature, i.e. it is the remnant of the magnetic field of parent molecular cloud core~\cite{dudorov95,fmft}. Possibility of dynamo generation of the magnetic field in the accretion discs is also considered in a number of works (see, for example, \cite{brandenburg95}, \cite{gressel15}, \cite{moss16}).

Accretion and differential rotation lead to generation and amplification of the magnetic field in the regions with high ionization fraction. Cosmic rays, X-rays and radioactive decay are the main sources of ionization in the ADYS. Attenuation of the cosmic and X-rays in the surface layers of the discs leads to formation of layered ioinization structure of the discs~\cite{gammie96}. Region of low ionization fraction, $x\leq 10^{-12}$, so-called `dead' zone, lies typically between 0.5-1~au and 10-30~au near the midplane of the accretion discs of T Tauri stars~\cite{sano00, mohanty13, fmfadys}. Ohmic and ambipolar diffusion hinder generation of the magnetic field inside the `dead' zones~\cite{fmfadys}. Dudorov and Khaibrakhmanov ~\cite{fmfadys} have shown that the magnetic field geometry varies over the disc. The magnetic field is frozen into gas and has quasi-azimuthal geometry in the inner region, where thermal ionization operates. The magnetic field preserves its initial poloidal geometry inside the `dead' zones. The magnetic field can have quasi-radial or quasi-azimuthal geometry in the outer regions of the ADYS depending on the intensity of ionization sources and dust grain parameters. 
We have not considered contribution of the dissipative MHD effects to thermal processes yet. Lizano et al.~\cite{lizano16} calculated thermal structure of the accretion discs taking into account turbulent and Ohmic heating, and found that both processes can have comparable effect on the temperature of the discs of T~Tauri stars. But they assumed that resistivity is uniform in the disc that does not allow to correctly study role of the Ohmic heating in the discs with `dead' zones.

In this work, our aim is to investigate effect of Ohmic and ambipolar heating on thermal structure of ADYS on the basis of our MHD model of the accretion disc~\cite{fmfadys, kh17}.

The paper is organized as follows. In section~\ref{Sec:problem}, we describe the problem statement and main equations of the model. Results of the simulations are presented in section~\ref{Sec:results}. We summarize and discuss the results in section~\ref{Sec:discuss}.

\section{Problem statement and main equations.}
\label{Sec:problem}
\begin{figure}
\begin{center}
\includegraphics[width=0.7\textwidth]{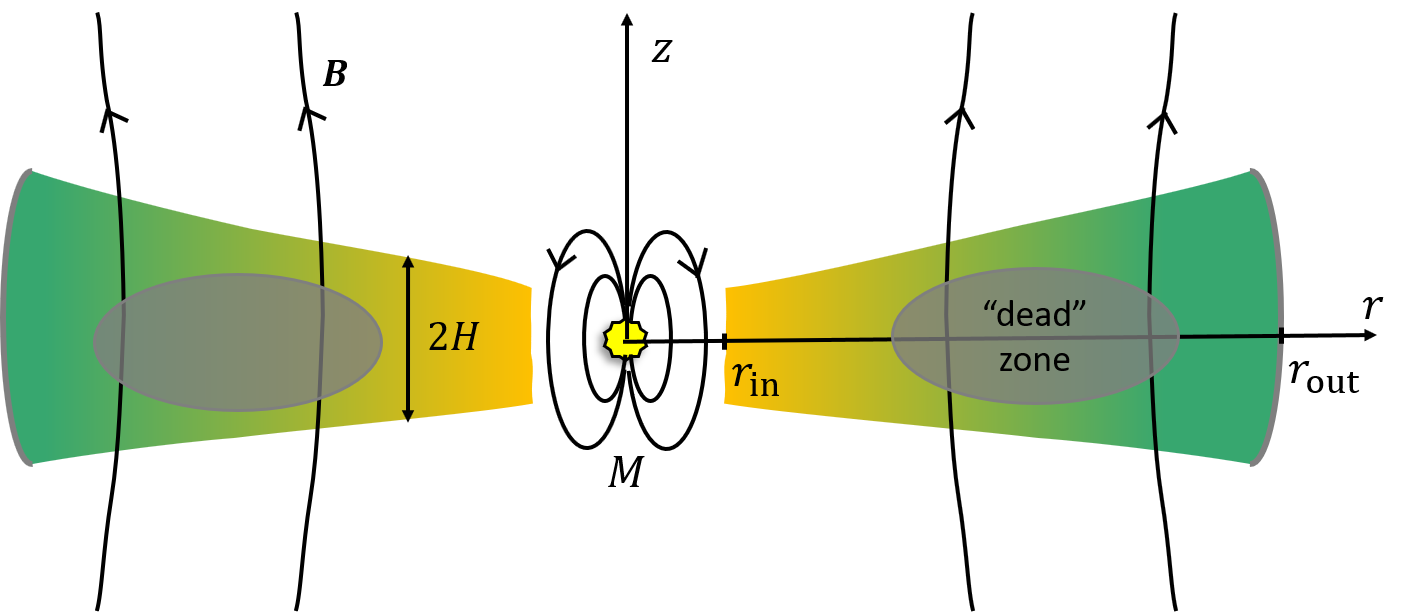}
\end{center}
\caption{Schematic view of the accretion disc with large scale magnetic field $\bvec{B}$ around star with mass $M$. The disc has inner boundary $r_{{\rm in}}$ (radius of stellar magnetosphere) and outer boundary $r_{{\rm out}}$ (contacts boundary with the interstellar medium), and has half-thickness $H$. Grey region inside the disc is the `dead' zone, that is the region of low ionization fraction.}
\label{Fig:scheme}
\end{figure}

We consider stationary geometrically thin and optically thick accretion disc around solar mass classical T Tauri star (see schematic picture in Figure~\ref{Fig:scheme}). We use cylindrical coordinates $(r,\,\varphi,\,z)$. Disc rotation axis is directed along $z$-axis. The disc is considered to be in hydrostatic equilibrium in the $z$-direction, so that the velocity components are $\bvec{v}=(v_r,\,v_{\varphi},\,0)$. Magnetic field components: $\bvec{B}=(B_r,\,B_{\varphi},\,B_z)$. Inner edge of the disc $r_{{\rm in}}$ is determined by the boundary of stellar magnetosphere. Outer edge $r_{{\rm out}}$ is the contact boundary with the interstellar medium.

We investigate the dynamics of the accretion discs using following system of MHD equations taking into account viscous stresses, thermal conductivity, magnetic ambipolar diffusion, Ohmic diffusion, and magnetic buoyancy:
\begin{eqnarray}
\frac{\partial\rho}{\partial t} + \Div\left(\rho\bvec{v}\right) &=& 0,\label{Eq:Contin}\\
\rho\frac{\partial\bvec{v}}{\partial t} + \left(\bvec{v}\nabla\right)\bvec{v}&=&-\Grad P - \rho \Grad\Phi +\frac{1}{4\pi}[\Rot\bvec{B},\,\bvec{B}] + \Div\sigma^{'},\\
\rho T\frac{\partial s}{\partial t} &=& \sigma_{ik}^{'}\frac{\partial v_i}{\partial x_k} + \Div\left(\kappa_{{\rm r}}\Grad T\right) + \Gamma,\label{Eq:EnergyMHD}\\
\frac{\partial \bvec{B}}{\partial t} &=& \Rot\left[\bvec{v} + \bvec{v}_{{\rm ad}},\,\bvec{B}\right] + \nu_{{\rm m}}\nabla^2\bvec{B} + \Rot\left[\bvec{v}_{{\rm b}},\,\bvec{B}_{{\rm t}}\right],\label{Eq:Induction}
\end{eqnarray}
where $\Phi$ is the gravitational potential of the star, $\sigma^{'}$ is the viscous stress tensor (see~\cite{LLVI}), $s$ is the entropy, $\kappa_{{\rm r}}$ is the radiative heat conductivity (see~\cite{mihalas_book}), $\Gamma$ describes additional heating sources, $\bvec{v}_{{\rm ad}}$ is the velocity of ambipolar diffusion, $\nu_{{\rm m}}$ is the magnetic viscosity, $\bvec{v}_{{\rm b}}$ is the buoyant velocity (see~\cite{dudorov89}), $\bvec{B}_{{\rm t}}=(0,\,B_{\varphi},\,0)$ is the toroidal component of the magnetic field,  $P=\rho R_{{\rm g}}T/\mu$, $s=c_{{\rm v}}{\rm ln}\left(P/\rho^{\gamma}\right)$, $R_{{\rm g}}$ is the universal gas constant, $\mu=2.3$ is the mean molecular weight of the gas, $c_{{\rm v}}$ is the specific heat at constant volume, $\gamma$ is the adiabatic index. Other quantities are used in their usual physical notation.

The first term on right-hand side of Equation (\ref{Eq:EnergyMHD}) describes heating of the gas by viscous friction, $\Gamma_{{\rm v}}$. In addition to the viscous heating, we consider heating by cosmic rays and radioactive decay, Ohmic (Joule) and ambipolar heating,
\begin{equation}
	\Gamma = \Gamma_{{\rm cr}} + \Gamma_{{\rm o}} + \Gamma_{{\rm a}}.
\end{equation}
Rate of heating by cosmic rays and radioactive decay $\Gamma_{{\rm cr}}$ is calculated according to \cite{dalessio98}. Rate of Ohmic heating
\begin{equation}
\Gamma_{{\rm o}} = \frac{\nu_{{\rm m}}}{4\pi}\left(\Rot\bvec{B}\right)^2. \label{Eq:Go}
\end{equation}
Rate of ambipolar heating~\cite{scalo77}
\begin{equation}
\Gamma_{{\rm a}} = \frac{1}{2}v_{{\rm ad}}^2R_{{\rm in}},\label{Eq:Ga}
\end{equation}
where ambipolar velocity in stationary approximation
\begin{equation}
\bvec{v}_{{\rm ad}} = \frac{[\Rot\bvec{B},\,\bvec{B}]}{4\pi R_{{\rm in}}}
\end{equation} 
and coefficient of friction between ions and neutrals
\begin{equation}
R_{{\rm in}} = \mu_{\rm{in}}n_{\rm{i}}n_{\rm{n}}\langle\sigma v\rangle_{\rm{in}},
\end{equation}
$\mu_{\rm{in}}$ is the reduced mass of ions and neutrals, $n_{\rm{i}}$ is the number density of ions, $n_{\rm{n}}$ is the number density of neutrals, $\langle\sigma v\rangle_{\rm{in}}=2\times 10^{-9}\,\mbox{cm}^3\,\mbox{s}^{-1}$ is the `slowing-down' coefficient~\cite{spitzer_book}. We consider plasma consisting of hydrogen, helium and metals with standard cosmic abundance. Mean mass of neutrals is $2.3\,m_{{\rm p}}$, mean mass of metal ions $m_{{\rm i}}=30\,m_{{\rm p}}$, where $m_{{\rm p}}$ is the proton mass.

We solve equations (\ref{Eq:Contin}-\ref{Eq:Induction}) in frame of Shakura and Sunyaev~\cite{ss73} approximation (see~\cite{fmfadys} and~\cite{kh17} for details). It is considered that angular momentum is transported radially by turbulent stresses. Molecular viscosity in the stress tensor $\sigma^{'}$ is replaced by turbulent viscosity $\nu_{{\rm t}} = \alpha v_{{\rm s}} H$, where $\alpha<1$ is the non-dimensional parameter characterizing turbulence efficiency, $v_{{\rm s}}$ is the sound speed, $H$ is the accretion disc's scale height. We assume that the disc is in centrifugal and hydrostatic balance, and that the heat released inside the disc is transported vertically by radiation. Radial derivatives are neglected in comparison with vertical derivatives, and the latter are replaced by finite differences. Then Equations (\ref{Eq:Contin}-\ref{Eq:EnergyMHD}) transform to
\begin{eqnarray}
  \dot{M} &=& - 2 \pi r v_r \Sigma, \label{Eq:Mass}\\
  \dot{M} \Omega_{{\rm k}} f &=& 2 \pi \alpha v_{{\rm s}}^2 \Sigma, \label{Eq:AngMom}\\
  v_{\varphi} &=& \sqrt{\frac{GM}{r}}\left(1 + \frac{z^2}{r^2}\right)^{-3/4},\\
  H &=& \frac{v_{{\rm s}}}{\Omega_{{\rm k}}},\\
  \sigma_{{\rm sb}} T_{{\rm eff}}^4 &=& \frac{3}{8\pi}\dot{M}\Omega_{{\rm k}}^2 f + 2H\Gamma_{{\rm cr}} + \nonumber\\
  & &  \frac{\nu_{{\rm m}}}{4\pi}\frac{B_r^2 + B_{\varphi}^2}{H} + \frac{\left(B_zB_r\right)^2 + \left(B_zB_{\varphi}\right)^2 + \left(B_r^2 + B_{\varphi}^2\right)^2}{32\pi^2 R_{{\rm in}}H},\label{Eq:Energy}\\
  T^4 &=& T_{{\rm eff}}^4 \left(\frac{1}{2} + \frac{3}{4}\tau\right) ,
\end{eqnarray}
where $\dot{M}$ is the accretion rate, $\Sigma=2\rho H$ is the surface density of the disc, $\Omega_{{\rm k}}=\sqrt{GM/r^3}$ is the Keplerian angular speed, $G$ is the gravitational constant, $f = 1 - \sqrt{r_{{\rm in}} / r}$,  $T_{{\rm eff}}$ is the effective temperature of the disc, $T$ is the temperature inside the disc, $\tau=\kappa\Sigma$ is the optical thickness, and $\kappa$ is the opacity. Solution of the induction Equation (\ref{Eq:Induction}) in approximations of our model:
  \begin{eqnarray}
  B_r &=& -\frac{v_r H}{\eta}B_z, \label{Eq:Br}\\
  B_{\varphi} &=& -\frac{3}{2}\left(\frac{H}{r}\right)^2\frac{v_{\varphi}H}{\eta}B_z - \frac{1}{2}\left(\frac{H}{r}\right)\frac{v_{\varphi}H}{\eta}B_r, \label{Eq:Bphi}\\
 B_z &=&
  \begin{cases}
  B_{z0}\frac{\Sigma}{\Sigma_0} & R_{{\rm m}} \gg 1,\\
  \sqrt{4\pi x\rho^2 r|v_r|} & R_{{\rm m}} \leq 1,
  \end{cases}\label{Eq:Bz}
\end{eqnarray}
where $B_{z0}$ and $\Sigma_0$ are the magnetic field strength and surface density at the outer edge of the disc, $x$ is the ionization fraction, $R_{{\rm m}}$ is the magnetic Reynolds number,
\begin{equation}
\eta = \nu_{{\rm m}} + \eta_{{\rm a}} + \eta_{{\rm b}}
\end{equation}
is the effective diffusivity including effects of Ohmic diffusion (the first term), ambipolar diffusion (the second term), and magnetic buoyancy (the third term). Magnetic viscosity 
\begin{equation}
	\nu_{{\rm m}} = \frac{c^2}{4\pi\sigma},
\end{equation}
where $\sigma$ is the electric conductivity (see book~\cite{alfven50}). Ambipolar diffusivity
\begin{equation}
\eta_{{\rm a}} =  \frac{B_z^2}{4\pi R_{{\rm in}}}.
\end{equation}
Diffusivity due to magnetic buoyancy
\begin{equation}
\eta_{{\rm b}} = H|v_{{\rm b}}|.
\end{equation}
Velocity of magnetic buoyancy $v_{{\rm b}}$ is determined from the balance
between buoyant and drag forces according to~\cite{kh17b}. Aerodynamic and turbulent drags are considered.

Solutions (\ref{Eq:Br}) and (\ref{Eq:Bphi}) determine radial and azimuthal components of the magnetic field from the balance between advection and diffusion in corresponding directions. Solution (\ref{Eq:Bz}) for $R_{{\rm m}} \gg 1$ describes the magnetic field that is frozen into gas in regions with high conductivity. Solution (\ref{Eq:Bz}) for $R_{{\rm m}} \leq 1$ determines $B_z$ from the equality of advection and ambipolar diffusion velocities in the regions with small conductivity.

Ionization fraction is calculated following~\cite{dud87}, $x=(n_{{\rm e}} + n_{{\rm i}}) / (n_{{\rm e}} + n_{{\rm i}} + n_{{\rm n}})$, where $n_{{\rm e}}\approx n_{{\rm i}}$ is the electrons concentration. Collisional ionization fraction $x_{{\rm s}}$ is calculated from the equation
\begin{equation}
\left(1 - x_{{\rm s}}\right)\xi = \alpha_{{\rm r}}x_{{\rm s}}^2n_{\rm{n}} + \alpha_{{\rm g}}x_{{\rm s}}n_{\rm{n}},
\end{equation}
where $\xi$ is the total ionization rate by cosmic rays, X-rays and decay of radionuclides, $\alpha_{{\rm r}}$ is the rate of radiative recombinations, $\alpha_{{\rm g}}$ is the rate of recombinations onto the dust grains. We consider spherical grains with initial mean radius $0.1\,\mu m$ and standard composition. Evaporation of grains is taken into account.

In the regions with high temperature $T>400$~K, we calculate thermal ionization $x_j^{{\rm T}}$ of hydrogen ($j={\rm H}$) and alkali metal atoms ($j={\rm m}$) using the Saha equation. `Mean' metal with ionization potential of $5.76$~eV and abundance logarithm of $5.97$ is considered for simplicity. Total ionization fraction
\begin{equation}
x = x_{{\rm s}} + \sum_j\nu_jx_j^{{\rm T}},\label{Eq:x}
\end{equation}
were $\nu_j$ is the abundance of element $j$ with respect to hydrogen.

System of algebraic equations (\ref{Eq:Mass}-\ref{Eq:x}) is non-linear. We solve it numerically using the iterative method together with the bisection method. Solution without Ohmic and ambipolar heating found by \cite{fmfadys} is used as an initial approximation.

\section{Results.}
\label{Sec:results}
In this work we present illustrative results for the following parameters: $M=1\,M_{\odot}$, stellar radius $R=2\,R_{\odot}$, magnetic field strength at stellar surface $B_{\star}=2$~kG, $\dot{M}=10^{-7}\,M_{\odot}\,{\rm yr}^{-1}$, $\alpha=0.01$, 
cosmic ray ionization rate $\xi_0=10^{-17}\,{\rm s}^{-1}$, cosmic ray attenuation length $100\,{\rm g}\,{\rm cm}^{-2}$. For simplicity, we use opacity $\kappa=10\,{\rm cm}^2\,{\rm g}^{-1}$ to approximately describe thermal structure of the disc at low temperatures. We set value of $B_{z0}$ at the outer boundary of the disc equal to the magnetic field strength of parent protostellar cloud (see details in~\cite{kh17}).

\subsection{Ionization fraction and magnetic field.}

\begin{figure}
\begin{center}
\includegraphics[width=0.8\textwidth]{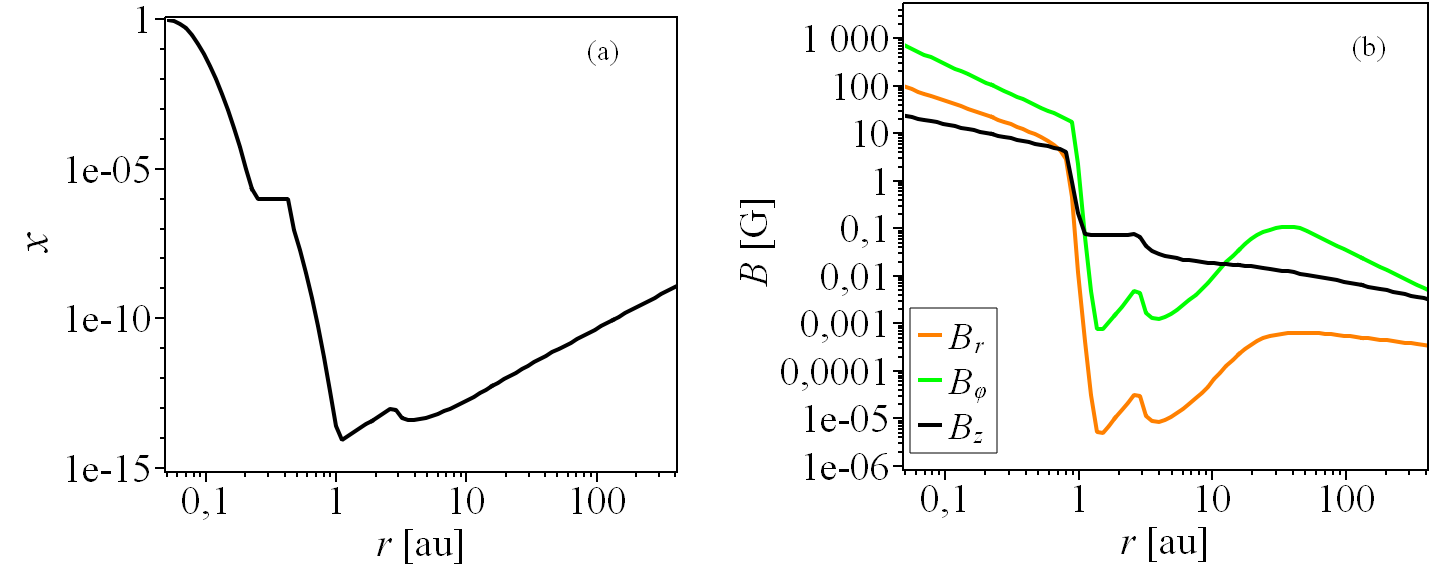}
\end{center}
\caption{Radial profiles of the ionization fraction in the midplane of the disc (panel (a)) and magnetic  field components at altitude $z=0.5\,H$ above the midplane (panel (b)).}
\label{Fig:xB}
\end{figure}

The Ohmic and ambipolar heating are determined by ionization fraction and magnetic field strength, as Equations (\ref{Eq:Go}) and (\ref{Eq:Ga}) show. In Figure~\ref{Fig:xB}, we plot radial profiles of the ionization fraction (panel (a)) and magnetic field components (panel (b)). It is considered that $B_{\varphi}$ strength  is limited by the equipartition value $B_{{\rm eq}}$. The equipartition value corresponds to plasma beta $\beta=1$.

Figure~\ref{Fig:xB}(a) shows that the radial profile of the ionization fraction is non-monotonic. The ionization fraction is high in the inner region of the disc, $r<1$~au, due to thermal ionization.  The ionization fraction increases with distance due to decrease of gas density and more efficient ionization by cosmic rays at $r>1$~au. There is region of low ionization fraction, $x\leq 10^{-12}$, at $0.8$~au $<r<$ $21$~au. This region is called `dead' zone~\cite{gammie96}. 

Figure~\ref{Fig:xB}(b) shows that there is three regions with different magnetic field geometry in the disc. In the region of thermal ionization, $r<1$~au, the magnetic field is frozen into gas, and efficient generation and amplification of all three components take place. Azimuthal component has fastest generation rate, and $B_{\varphi}>(B_r,\,B_z)$ in this region. Inside the `dead' zone, the magnetic field is quasi poloidal due to efficient Ohmic diffusion, $B_z\gg (B_r,\,B_{\varphi})$. In the outer region, $r>20$~au, ambipolar diffusion hinders generation of $B_r$, and the magnetic field has quasi-azimuthal geometry, $B_{\varphi} \geq B_z$.

\subsection{Effect of Ohmic and ambipolar heating  on thermal structure of the disc.}

\begin{figure}
\begin{center}
\includegraphics[width=0.8\textwidth]{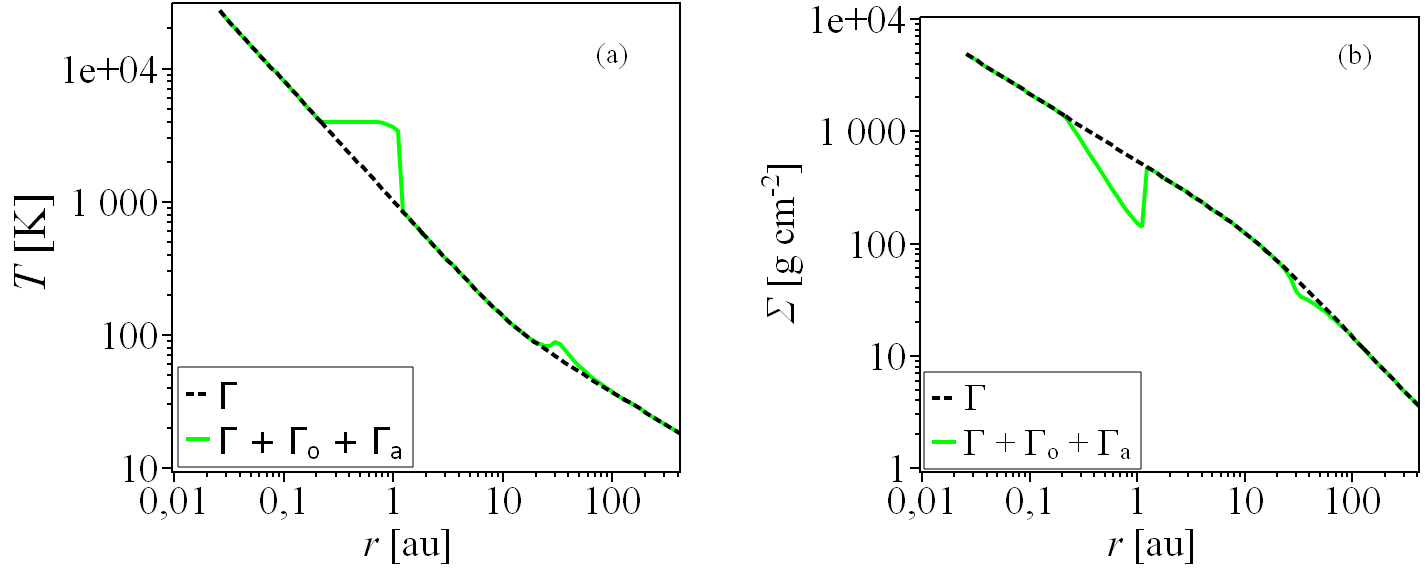}
\end{center}
\caption{Radial profiles of temperature in the midplane of the disc (panel (a)) and surface density of the disc (panel (b)). Dashed lines: simulation taking into account viscous and cosmic ray heating, green solid line: simulation with additionally included Ohmic and ambipolar heating.}
\label{Fig:TS}
\end{figure}

As it was mentioned in section~\ref{Sec:problem}, four heating sources are considered in our model of the disc: viscous heating ($\Gamma_{{\rm v}}$), heating by cosmic rays and radioactive decay ($\Gamma_{{\rm cr}}$), Ohmic ($\Gamma_{{\rm o}}$) and ambipolar heating ($\Gamma_{{\rm a}}$). 
We carried out two simulations: one with viscous and heating by cosmic rays and radioactive decay ($\Gamma_{{\rm v}} + \Gamma_{{\rm cr}}$), and another with additional effect of Ohmic and ambipolar heating ($\Gamma_{{\rm v}} + \Gamma_{{\rm cr}} + \Gamma_{{\rm o}}+\Gamma_{{\rm a}}$). In Figure~\ref{Fig:TS}, we plot radial profiles of disc temperature (panel (a)) and surface density (panel (b)) for this two cases. Figure~\ref{Fig:TS}(a) shows that Ohmic and ambipolar heating leads to increase of the temperature near the edges of the `dead' zone. Temperature increases by $1000-3000$~K near the inner edge, $r\approx (0.5-1)$~au, and by $100$~K near the outer edge, $r\approx (30-50)$~au. Figure~\ref{Fig:TS}(b) shows that temperature growth leads to decrease in surface density. It happens because the accretion speed $v_r$ is higher in regions with high temperature, and therefore matter is accreted more efficiently from these regions.

Temperature growth and density decrease lead to increase of the ionization fraction in the regions of Ohmic and ambipolar heating. As a consequence, size of the `dead' zone reduces. The `dead' zone is situated between $1.3$~au and $20$~au, when Ohmic and ambipolar heating operate.

\section{Conclusions.}
\label{Sec:discuss}
We investigated thermal structure of the magnetized accretion discs of classical T Tauri stars. Ohmic and ambipolar heating are included in the MHD model of the accretion discs.

The simulations have shown that the dissipative MHD effects operate near the boundary of the `dead' zone. The `dead' zone is situated between $0.8$~au and $21$~au in the simulation without Ohmic and ambipolar heating. Ambipolar heating is determined by $B_{\varphi}$ strength in the regions of maximal ambipolar velocity. In the case of maximum $B_{\varphi}=B_{{\rm eq}}$, temperature increases by value $\leq 3000$~K near the inner boundary of the `dead' zone, $r\approx (0.5-1)$~au, and by 100~K near its outer boundary $r\approx (30-50)$~au. If the magnetic buoyancy operates, then $B_{\varphi}\approx B_z$,  and ambipolar heating affects only inner boundary of the `dead' zone.

Radial extent of the `dead zone' reduces due to the action of the Ohmic and ambipolar heating. We suppose that size of the `dead' zone in the vertical direction would also reduce. Amplification of the fossil magnetic field is possible outside the `dead' zones, as well as dynamo generation from the seed fossil field. We conclude that the dissipative MHD effects support amplification and generation of the magnetic field in the accretion discs of young stars.

Temperature growth due to magnetic field dissipation can lead to evaporation of dust grains near the borders of `dead' zones in the accretion discs of young stars. These regions with evaporated dust grains can probably be observed in the maps of discs' thermal radiation.

\Thanks{We thank anonymous referee for useful comments that helped to improve our paper. The work is supported by Russian Foundation for Basic Research (project 18-02-01067) and by the Ministry of Science and High Education (the basic part of the State assignment, RK no. AAAA-A17-
117030310283-7). }



 \bibliographystyle{mhd}
 \bibliography{khaibrakhmanov_bib}


\lastpageno	


\end{document}